\documentclass[10pt,twocolumn]{article}

\usepackage[letterpaper,
            top=57pt, bottom=73pt,
            left=54pt, right=54pt,
            columnsep=17pt]{geometry}

\usepackage[T1]{fontenc}
\usepackage{lmodern}
\usepackage{microtype}

\usepackage{amsmath,amssymb,amsthm}
\usepackage{bbm}

\usepackage{algorithm}
\usepackage{etoolbox}
\AtBeginEnvironment{algorithm}{\color{black}}
\usepackage{algpseudocode}

\usepackage{graphicx}
\usepackage{svg}

\usepackage{xcolor}
\definecolor{acmblue}{RGB}{0,80,158}

\usepackage{titlesec}
\titleformat{\section}{\normalfont\large\bfseries\scshape}{\thesection.}{0.5em}{}
\titleformat{\subsection}{\normalfont\normalsize\bfseries}{\thesubsection.}{0.5em}{}
\titleformat{\subsubsection}{\normalfont\normalsize\itshape}{\thesubsubsection.}{0.5em}{}
\titlespacing*{\section}{0pt}{8pt plus 2pt minus 2pt}{4pt plus 1pt}
\titlespacing*{\subsection}{0pt}{6pt plus 2pt minus 2pt}{3pt plus 1pt}

\usepackage{abstract}

\usepackage[font=small,labelfont=bf,labelsep=period]{caption}

\usepackage{booktabs}
\usepackage{float}

\usepackage[colorlinks=true,
            linkcolor=acmblue,
            citecolor=acmblue,
            urlcolor=acmblue]{hyperref}

\usepackage[backend=biber,
            style=numeric-comp,
            sorting=none,
            giveninits=true,
            maxbibnames=99]{biblatex}
\usepackage{fancyhdr}
\fancypagestyle{firstpage}{%
  \fancyhf{}
  
  \fancyfoot[C]{\small\thepage}
}

\usepackage{authblk}

\newcommand{\ccsdesc}[2][100]{%
  \ifnum#1>499 \textbf{#2}\else
  \ifnum#1>299 \textit{#2}\else
  #2\fi\fi}

\newcommand{\anontext}[2]{#2}

\begin{document}

\title{\vspace{-1em}\LARGE\bfseries
  FPLIER: Federated Pathway-Level Information Extractor}

\author[1,2]{Daniele Malpetti}
\author[1]{Christian Berchtold}
\author[1,2]{Francesco Gualdi}
\author[1]{Marco Scutari}
\author[1,2$\dagger$]{Laura Azzimonti}
\author[1,2$\dagger$]{Francesca Mangili}

\affil[1]{Dalle Molle Institute for Artificial Intelligence (IDSIA),
          USI-SUPSI, Lugano, Switzerland}
\affil[2]{Swiss Institute of Bioinformatics (SIB), Lugano, Switzerland}
\affil[ ]{{\small $^\dagger$Both authors contributed equally to this research.}}
\affil[ ]{{\small\texttt{\{daniele.malpetti, christian.berchtold,
           francesco.gualdi, marco.scutari, laura.azzimonti,
           francesca.mangili\}@supsi.ch}}}

\date{%
  Accepted for publication at ACM BCB '26, Rende (CS), Italy}

\maketitle
\thispagestyle{firstpage}

\begin{abstract}
In transcriptomics, gene-set-aware factorization methods such as the Pathway Level Information Extractor (PLIER) are most effective when trained on large, heterogeneous expression compendia. Yet, many clinically relevant cohorts cannot be pooled into a single dataset due to privacy and governance constraints. We present \emph{FPLIER}, a federated extension of PLIER that enables distributed training across multiple data holders while incorporating publicly available datasets. Through secure aggregation, FPLIER produces training updates algebraically equivalent to those of a centralized pooled-data approach while keeping expression data local. We evaluate FPLIER across multiple scenarios in two simulated consortia (from the K--CLIER and MultiPLIER studies) and demonstrate stable convergence. We further conduct a systematic analysis of membership inference attacks targeting both intermediate training statistics and the released model. 
Our results show that privacy risk is governed by the rank of the training expression matrix. Incorporating public data or reducing data dimensionality increases this rank, moving the system toward a full-rank regime in which training and non-training samples become indistinguishable to the attacker, and membership-inference performance approaches random guessing.
\end{abstract}




\section{Introduction}

Large-scale transcriptomic profiling has become a cornerstone of modern biomedical research, enabling the systematic investigation of biological processes across diseases, tissues, and experimental conditions, from bulk RNA sequencing \cite{wang2009rna} to single-cell transcriptomics \cite{stubbington2017single}. Nevertheless, transcriptomic analysis poses significant methodological challenges because gene expression data are high-dimensional and many studies involve limited sample sizes \cite{clarke2008properties}, particularly in rare-disease and translational settings.

To address these challenges, gene-set-aware methods are widely employed to incorporate prior biological knowledge into transcriptomic analysis, thereby reducing effective dimensionality and enhancing interpretability. These approaches range from enrichment-based strategies, such as single-sample gene set scoring methods \cite{barbie2009systematic, hanzelmann2013gsva}, to latent variable (LV) models in which gene sets guide the structure of the learned components. 
Among them, the Pathway Level Information Extractor (PLIER)~\cite{mao2019pathway} combines matrix factorization with gene set constraints to learn LVs, encouraging a subset of these LVs to align with the specified gene sets.
By shifting analysis from individual genes to biologically meaningful LVs, PLIER enables robust downstream analyses such as differential expression and predictive modeling, while improving interpretability and mitigating noise and multiple-testing effects.

Prior work indicates that PLIER models benefit substantially from training on large, heterogeneous expression compendia, which help produce latent factors that are both robust and well disentangled \cite{taroni2019multiplier}. In addition, incorporating data from biologically relevant contexts, such as disease-specific cohorts, can further refine latent representations and improve downstream utility~\cite{taroni2019multiplier}. Although large public repositories now host tens of thousands of bulk RNA-seq samples~\cite{collado2017reproducible, tcga}, many domain-specific datasets remain within individual laboratories, hospitals, and research consortia. These non-public datasets are often constrained by privacy concerns, regulatory requirements, and data governance policies that limit centralized aggregation and require data to remain local.

Federated learning (FL)~\cite{mcmahan2016communication} offers a natural framework for preserving institutional data control and data locality by enabling collaborative model training across multiple data holders without pooling the underlying data. In this paradigm, each participating institution performs computations locally on its own dataset and shares only selected intermediate quantities with a coordinating server, which aggregates them to update a global model. FL has attracted increasing interest in bioinformatics, where data-sharing restrictions are common across many domains, and a range of federated methods have recently been proposed for various analytical tasks~\cite{malpetti2025technical}.

Motivated by these considerations, we propose \emph{FPLIER}, a federated extension of PLIER. To our knowledge, FPLIER is the first federated formulation of a gene-set-aware factorization method for transcriptomic analysis. In contrast to deep learning models, for which well-established, generally applicable federated training strategies exist, gene-set-aware factorization methods do not naturally align with standard FL workflows. Consequently, their federated implementation requires a dedicated reformulation of the underlying optimization procedure.

Specifically, this work makes the following contributions.
\begin{itemize}
  \item We introduce FPLIER, a federated reformulation of PLIER that enables collaborative training across multiple data holders without sharing raw gene expression data, while leveraging public transcriptomic repositories.
  \item We empirically validate FPLIER across simulated federated consortia derived from large RNA-seq compendia, demonstrating stable convergence.
  \item We systematically evaluate the robustness of the FPLIER protocol to membership inference attacks during training and after model release, showing that the rank of the gene expression training data governs attack effectiveness. The postrelease analysis also applies to standard centrally trained PLIER models and, to the best of our knowledge, is presented here for the first time.
\end{itemize}

The remainder of the paper is organized as follows.
Section~\ref{sec:background} reviews relevant background on PLIER and federated learning,
Section~\ref{sec:methods} introduces the FPLIER framework,
Section~\ref{sec:experiments} describes the experimental evaluation,
Section~\ref{discussion} discusses our findings,
and Section~\ref{sec:conclusions} concludes the paper.

\section{Background} \label{sec:background}

We review PLIER and federated learning (FL), and outline relevant adversarial attacks and privacy-enhancing technologies.

\begin{figure*}[t]%
\center \includegraphics[width=.9\textwidth]{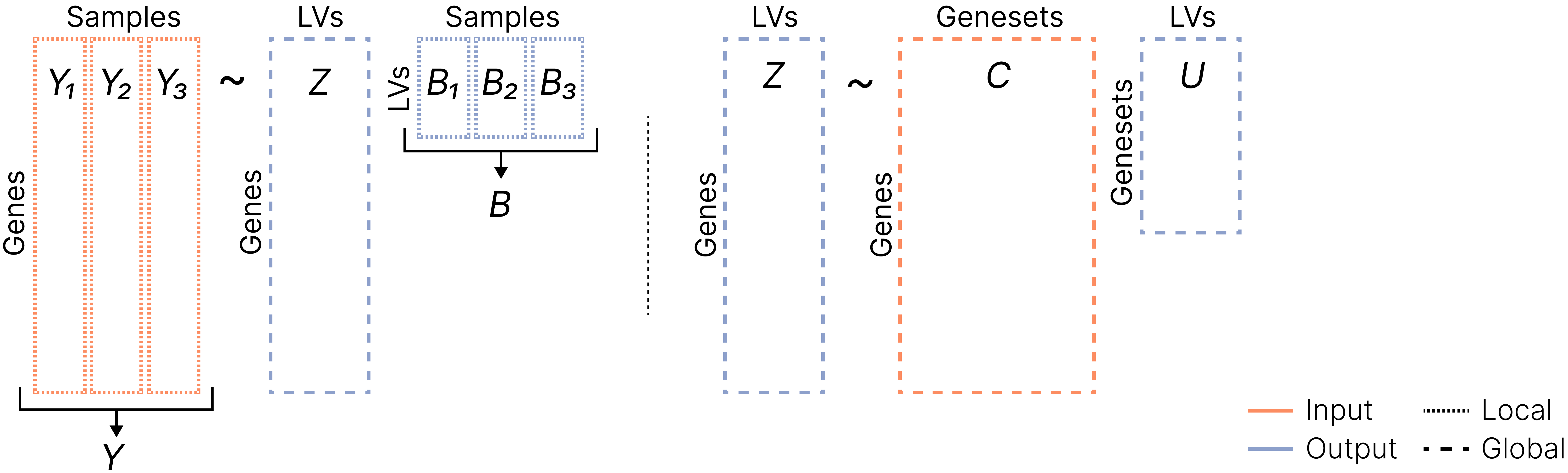}
\caption{Overview of the PLIER and FPLIER decompositions. PLIER takes as input a gene expression matrix $Y$ and a gene-set membership matrix $C$ encoding prior biological knowledge. The method learns matrices $Z$, $U$, and $B$ such that $Y \approx ZB$ and $Z \approx CU$. Here, $B$ provides a low-dimensional representation of samples in terms of latent variables (LVs), $Z$ contains the corresponding gene loadings, and $U$ links LVs to gene sets. In the federated setting (FPLIER), $Y$ and $B$ are partitioned column-wise across clients, with each client storing and processing its own subset of samples locally, whereas $Z$, $U$, and $C$ are global and visible to all consortium members.}
\label{fig.scheme_2}
\end{figure*}

\subsection{PLIER} \label{sec:plier}

The Pathway-Level Information Extractor (PLIER)~\cite{mao2019pathway} is a matrix factorization method designed to analyze bulk transcriptomic data. Its objective is to derive a low-dimensional representation of each sample, not in terms of individual genes, but in terms of latent variables (LVs), a subset of which can be directly interpreted using prior biological knowledge. This prior knowledge is encoded through selected gene sets, such as biological pathways or single-cell–derived signatures. When single-cell signatures are used, PLIER can be interpreted as providing an approximate deconvolution of bulk expression profiles into contributions associated with specific cell types or cell states~\cite{malpetti2025protocol}.

PLIER has been applied across diverse biological contexts, demonstrating both versatility and practical impact. A prominent example is the study of rare diseases by Taroni et al.~\cite{taroni2019multiplier}, who introduced MultiPLIER, a transfer-learning framework in which a PLIER model trained on a large, heterogeneous public compendium spanning multiple tissues and conditions is transferred to small datasets, enabling the extraction of coordinated, biologically interpretable expression patterns and improving the characterization of disease-related processes relative to models trained on individual studies. Another application is the K-CLIER model~\cite{clier}, a PLIER-based transformation tailored to kidney biology that integrates single-cell–derived prior information to extract interpretable signals associated with kidney cell types and states, and has been used to study the role of proximal tubules in acute kidney injury. Beyond human biology, the MousiPLIER framework~\cite{zhang2024mousiplier} extends PLIER to mouse transcriptomic data by training a large-scale model on over 190,000 mouse brain RNA-seq samples, yielding pathway- and cell type–aligned LVs that enhance interpretability and reveal biologically meaningful aging-associated signals in brain cell populations.

Formally, given a gene expression matrix $Y \in \mathbb{R}^{p \times n}$ with $p$ genes and $n$ samples, PLIER incorporates prior biological knowledge through a binary gene set membership matrix $C \in \{0,1\}^{p \times m}$, where $C_{gm} = 1$ indicates that gene $g$ belongs to gene set $m$. Using this structure, PLIER learns a factorization of the form
\[
Y \approx ZB, \qquad Z \approx CU,
\]
where $Z \in \mathbb{R}^{p \times k}$ contains gene loadings for the $k$ latent variables, $B \in \mathbb{R}^{k \times n}$ represents samples in the latent space, and $U \in \mathbb{R}^{m \times k}$ encodes the associations between latent variables and gene sets. An overview of the PLIER decomposition and the role of the individual matrices is shown in Figure~\ref{fig.scheme_2}A.

The matrices $Z$, $B$, and $U$ are then learned during training. Conceptually, $Z$ captures gene-level contributions to each LV, while $B$ provides a compact representation of each sample in the latent space. Biological interpretability arises through $U$, which links LVs to known gene sets and enables selected components of $B$ to be interpreted in terms of established biological processes.

Formally, PLIER estimates the matrices by solving the optimization problem
\begin{equation}
\min_{Z,B,U} \;\; 
\|Y - ZB\|_F^2
+ \lambda_1 \|Z - CU\|_F^2
+ \lambda_2 \|B\|_F^2
+ \lambda_3 \|U\|_1 ,
\label{eq:plier_obj}
\end{equation}
where $\|\cdot\|_F$ denotes the Frobenius norm and $\|\cdot\|_1$ the element-wise $\ell_1$ norm, and $\lambda_1,\lambda_2,\lambda_3 > 0$ are regularization parameters controlling the strength of the respective penalties. The first term enforces accurate reconstruction of the expression matrix; the second encourages gene loadings to align with the gene set structure encoded in $C$; and the remaining terms provide regularization.

PLIER training begins by estimating the hyperparameters $\lambda_1, \lambda_2$, and the number of LVs to use. These quantities are derived from a singular value decomposition (SVD) of the expression matrix $Y$.
The optimization problem is then solved using an alternating optimization scheme where each matrix is updated while the others are held fixed. At each iteration $\ell$, the matrices are updated as
\begin{equation} \label{eq:update}
\begin{aligned} 
Z^{(\ell+1)} &\gets
\big(Y {B^{(\ell)}}^\top + \lambda_1 C U^{(\ell)}\big)
\big(B^{(\ell)} {B^{(\ell)}}^\top + \lambda_1 I\big)^{-1} \,, \\
U^{(\ell+1)} &\gets
\arg\min_U \|Z^{(\ell+1)} - C U\|_F^2 + \lambda_3 \|U\|_1 \,, \\
B^{(\ell+1)} &\gets
\big({Z^{(\ell+1)}}^\top Z^{(\ell+1)} + \lambda_2 I\big)^{-1}
{Z^{(\ell+1)}}^\top Y \,.    
\end{aligned}
\end{equation}

In contrast to unconstrained factorizations such as SVD, PLIER incorporates structural constraints that guide the latent space toward biologically meaningful patterns. As a result, the representation learned by PLIER is both low-dimensional and partially interpretable. To keep the main text focused, we defer the full description of the optimization procedure, including detailed pseudocode, to Appendix~\ref{appendix:plier}.

\subsection{Federated Learning}

Federated learning (FL)~\cite{mcmahan2016communication} is a distributed machine learning paradigm in which multiple parties, commonly referred to as \emph{clients}, collaboratively train a shared \emph{global model} while keeping their data local. A central server maintains the global model and orchestrates the training process. In the standard deep-learning-oriented formulation~\cite{mcmahan2016communication}, each communication round typically proceeds as follows: the server broadcasts the current global model to a subset of clients; each selected client performs a small number of local optimization steps on its own data; the client then returns only the resulting parameter updates, or \emph{local model}, to the server.

Upon receiving the local updates, the server aggregates them to obtain an improved global model, often through a weighted averaging scheme such as FedAvg~\cite{mcmahan2016communication}. This updated model is then redistributed to clients for the next round. The process repeats until a convergence criterion is met, producing a single model that incorporates information from all participating datasets without ever requiring centralization of the underlying records.

In practice, FL has been developed and applied primarily to deep learning models, for which gradient-based optimization and parameter averaging fit naturally into this workflow. In contrast, the method considered in this work, PLIER, is a matrix factorization approach with closed-form linear-algebraic updates. Consequently, it does not map directly onto the standard FL training loop and requires a dedicated reformulation of its optimization procedure.

\subsubsection{Privacy-enhanced federated learning}

Although FL avoids the explicit centralization of raw data, data locality alone is not sufficient to ensure privacy. The primary privacy risks in FL arise from \emph{inference attacks} such as membership inference and model inversion~\cite{Zhou_2024}. Membership inference attacks aim to determine whether a specific individual’s data were included in the training process. In contrast, model inversion attacks attempt to reconstruct sensitive properties of training samples from shared model parameters or intermediate quantities. Such attacks may occur during training, if client-level updates are exposed, or after training, if the final model is publicly released in a white-box setting.

To mitigate these risks, practical FL systems often incorporate additional privacy-enhancing mechanisms beyond data locality. One widely adopted approach is \emph{secure aggregation}, which ensures that the server can access only aggregated information across clients and never individual client contributions. Conceptually, secure aggregation enables multiple participants to jointly compute a global sum of local updates while keeping each individual update private. As a result, the server observes only the final aggregate, substantially limiting, though not eliminating, the information that can be inferred about any single participant. An example of such a protocol is SecAgg+~\cite{bell}, which is designed to operate efficiently at scale, with communication costs increasing linearly in the dimensionality of the aggregated updates~\cite{li21}.

Another widely employed method is Differential Privacy (DP)~\cite{dwork2014algorithmic}, which provides a rigorous mathematical framework ensuring that the contribution of any single individual’s data has only a limited effect on the outcome of a computation. Intuitively, a differentially private algorithm produces outputs that are nearly indistinguishable whether or not a particular record is included in the dataset, thereby bounding what can be inferred about any individual. In practice, DP is typically achieved by adding carefully calibrated noise to computations or model updates. The goal is to enable useful statistical analysis while ensuring that participation does not substantially increase the risk of disclosing private information. However, achieving strong privacy guarantees often requires injecting noise at levels that may degrade model utility, particularly in high-dimensional biomedical settings.

Secure aggregation and DP are complementary and address different threat surfaces. Secure aggregation prevents the server from observing individual client updates, whereas DP provides formal guarantees on the information that can be inferred from aggregated statistics or released models. In this work, we rely on secure aggregation and empirically identify regimes that appear sufficiently safe in practice. Incorporating DP would introduce an explicit utility--privacy trade-off, which we leave for future work.

\subsubsection{Federated learning in bioinformatics}

In bioinformatics, federated formulations have been proposed across multiple molecular domains~\cite{malpetti2025technical}, including genomics, transcriptomics, and proteomics. In genomics, several applications focus on genome-wide association studies (GWAS), in which genomic variants are associated with phenotypes, typically using regression models~\cite{uffelmann2021genome}. In this setting, the privacy-preserving nature of FL is particularly valuable given the high sensitivity of genotype data. For example, Li et al.~\cite{li2022federated} developed federated approximations of generalized linear mixed-effects models that account for population structure through a genomic relatedness matrix. In transcriptomics, federated approaches have been explored for both bulk and single-cell RNA-seq data, including methods for differential gene expression and cell-type classification using both deep and non-deep learning models~\cite{sav2022privacy,wang2024scfed,zolotareva2021flimma}. Finally, in proteomics, a representative example is ProCanFDL~\cite{cai2025federated}, a privacy-preserving cancer subtyping model that integrates data from multiple cohorts and platforms while achieving performance comparable to that of centralized training.

\section{Methods} \label{sec:methods}

In this section, we describe the FPLIER training protocol and the membership-inference attacks used to evaluate its privacy risk.

\subsection{FPLIER protocol} 

The FPLIER protocol consists of three phases: (i) preliminary steps, (ii) federated training, and (iii) an optional privacy evaluation.  
An overview of all phases is provided in Algorithm~\ref{alg:realworld_protocol}. FPLIER supports training on a combination of private, client-specific gene expression datasets $\{Y_i^{\mathrm{priv}}\}_{i=1}^N$, where $i = 1, \dots, N$ indexes the participating clients, together with a shared public gene expression dataset $Y^{\mathrm{pub}}$.
As discussed in the following sections, incorporating public data is key to mitigating the method’s membership-inference risk.

We implemented FPLIER using the Flower FL framework~\cite{flower}. In a recent comparative analysis of open-source FL frameworks, Riedel et al.~\cite{riedel} identified Flower as one of the most mature and capable platforms based on criteria such as extensibility, usability, and community support. Our implementation uses Flower's built-in support for secure aggregation via the SecAgg+ protocol. The results in this work are obtained using Flower’s simulation engine; deployment in a real-world distributed consortium would requireonly configuring network endpoints and infrastructure, with no modifications to the Python implementation of the method.

\begin{algorithm}[t]
\small
\caption{FPLIER protocol}
\label{alg:realworld_protocol}
\begin{algorithmic}[1]
\Require Gene-set matrix $C$; public gene expression data $Y^{\mathrm{pub}}$; private gene expression data $\{Y_i^{\mathrm{priv}}\}_{i=1}^N$; train/holdout ratio $\rho$
\Ensure Final model $(C,Z,U,\lambda_2)$

\Statex \hspace{-\algorithmicindent}\textit{Phase 1: preliminary steps}
\State \textbf{Server:} Partition $Y^{\mathrm{pub}}$ into $\{Y^{\mathrm{pub}}_i\}_{i=1}^N$ and send $Y^{\mathrm{pub}}_i$ to client $i$
\ForAll{$i \in \{1,\dots,N\}$ \textbf{in parallel}}
\State \textbf{Client $i$:} $Y_i^{\text{local}} \gets Y_i^{\mathrm{priv}} \cup Y^{\mathrm{pub}}_i$
\State \textbf{Client $i$:} split $Y_i^{\text{local}}$ into $(Y_i,\,Y_i^{\text{holdout}})$ with ratio $\rho$
\EndFor

\State \textbf{Consortium:} z-score standardize $\{Y_i\}_{i=1}^N$ (Algorithm~\ref{alg:fed_standardization})

\Statex \hspace{-\algorithmicindent}\textit{Phase 2: federated training}
\State \textbf{Consortium:} Train FPLIER on $\{Y_i\}_{i=1}^N$ (Algorithm~\ref{alg:fed})
\State \textbf{Server:} Output final model $(C,Z,U,\lambda_2)$

\Statex \hspace{-\algorithmicindent}\textit{Phase 3: postrelease membership inference risk assessment (optional)}
\ForAll{$i \in \{1,\dots,N\}$ \textbf{in parallel}}
  \ForAll{$y_{ij} \in Y_i^{\text{priv}}$}
  \State \textbf{Client $i$:} $b_{ij} \gets (Z^\top Z + \lambda_2 I)^{-1} Z^\top y_{ij}$
  \State \textbf{Client $i$:} $\hat{y}_{ij} \gets Z b_{ij}$
  \State \textbf{Client $i$:} $s_{\text{release}}(y_{ij}) \gets \rho_{\mathrm{S}}(y_{ij}, \hat{y}_{ij})$
  \State \textbf{Client $i$:} $\ell_{ij} \gets 1$ if $y_{ij} \in Y_i$, $0$ if $y_{ij} \in Y_i^{\text{holdout}}$
  \EndFor
  \State \textbf{Client $i$:} Compute AUROC from  $s_{\text{release}}(y_{ij})$ and  $\ell_{ij}$
  \State \textbf{Client $i$:} If privacy risk unacceptable, notify server
\EndFor
\end{algorithmic}
\end{algorithm}

\begin{table*}[t]
\center
\small
 \caption{Overview of private and public cohorts used in the experimental design for the K--CLIER and MultiPLIER settings.}
  \label{tab:cohorts}
  \begin{tabular}{llll}
    \toprule
    \textbf{Setting} & \textbf{Cohort / study} & \textbf{Role} & \textbf{Samples} \\
    \midrule
    \textbf{K--CLIER}
      & KIRC (Kidney Renal Clear Cell Carcinoma) & Private cohort 1 & 618 \\
      & KIRP (Kidney Renal Papillary Cell Carcinoma) & Private cohort 2 & 323 \\
      & KICH (Kidney Chromophobe) & Private cohort 3 & 91 \\
      & Remaining TCGA samples & Shared public cohort & 10{,}316 \\
    \midrule
    \textbf{MultiPLIER}
      & SRP048759 (Leucegene AML sequencing) & Private cohort 1 & 430 \\
      & SRP050223 (T cell acute lymphoblastic leukemia) & Private cohort 2 & 400 \\
      & SRP050000 (Sepsis survival vs.\ death) & Private cohort 3 & 129 \\
      & SRP049820 (Endotoxin tolerance / sepsis risk) & Private cohort 4 & 83 \\
      & Remaining recount2 samples & Shared public cohort & 35{,}990 \\
    \bottomrule
  \end{tabular}
\end{table*}

\subsubsection{Phase 1: Preliminary steps}

The server first partitions the public expression matrix $Y^{\text{pub}}$ into $N$ shards of approximately equal size, denoted $\{Y_i^{\text{pub}}\}_{i=1}^N$, and distributes one shard to each client. This step incorporates public data as an added layer of protection against server‑side inference attacks, as further discussed in Section \ref{sec:privacy_discussion}.
Each client then constructs its local expression matrix as
\[
Y_i^{\text{local}} = Y_i^{\text{priv}} \cup Y_i^{\text{pub}},
\]
where the union denotes column-wise concatenation of samples. Although $Y_i^{\text{local}}$ includes public data, it remains entirely local to the client throughout the protocol.

Each local matrix $Y_i^{\text{local}}$ is then split into training and holdout subsets $(Y_i, Y_i^{\text{holdout}})$ using a predefined ratio. The training subset is used for federated model fitting, while the holdout subset is kept for the optional postrelease membership inference assessment. Both the allocation of samples to clients and the train/holdout split are performed locally and not disclosed to other consortium members; this uncertainty contributes importantly to the overall security of the procedure, as discussed in Section~\ref{discussion}.

Before training, in line with common practice in PLIER workflows, the consortium performs a federated gene-wise $z$-score standardization of the expression matrix. This step requires computing the global mean and variance of each gene across all participating datasets. These quantities are aggregated gene-level summary statistics that do not reveal per-sample expression values and are computed in a federated manner using secure aggregation. Additional implementation details are provided in Appendix~\ref{appendix:standardization}.

\subsubsection{Phase 2: Federated training}

During federated training, each client retains the expression matrix $Y_i$ and the corresponding latent representation $B_i$ locally, and never shares them with the server or other clients. In contrast, $Z$, $C$, and $U$ are global model parameters shared between the consortium. Figure~\ref{fig.scheme_2}B provides a visual overview of the FPLIER decomposition and highlights its contrast to centralized PLIER. A detailed description of the federated implementation is provided in Appendix~\ref{appendix:training}.

At the start of training, the consortium jointly estimates hyperparameters and latent dimensionality from the distributed matrix
\begin{equation} \label{eq:yyt}
    Y Y^\top 
    = \sum_{i=1}^N Y_i Y_i^\top \,.
\end{equation}
The server computes the eigendecomposition of this aggregated matrix (see Appendix~\ref{appendix:privacy} for the relationship between the decompositions of $Y$ and $YY^\top$) to determine the latent dimensionality $k$ and the regularization parameter $\lambda_2$, which are then communicated to the clients. Each client then initializes its matrix $B_i$ locally using its expression matrix $Y_i$.

The algorithm then enters an iterative phase that minimizes the objective in Eq.~\eqref{eq:plier_obj} in a distributed manner, following the update rules in Eq.~\eqref{eq:update}. In the federated setting, this requires computing at each iteration the sums of local matrix products
\begin{equation} \label{eq:bbtybt}
Y B^\top 
= \sum_{i=1}^N Y_i B_i^\top, \quad
B B^\top 
= \sum_{i=1}^N B_i B_i^\top .
\end{equation}

Because the quantities in Eqs.~\eqref{eq:yyt} and \eqref{eq:bbtybt} decompose additively across clients, they can be computed using the SecAgg+ protocol. Since the PLIER updates depend on the data only through these aggregated matrix products, replacing centralized computations with securely aggregated sums preserves the algebraic form of the centralized updates, up to minor numerical differences introduced by the SecAgg+ protocol (e.g., due to floating-point rounding). If the same initialization were used, the resulting iterates would coincide with those of the centralized PLIER. In our implementation, initialization is performed locally on each client, so the training trajectory need not match the centralized one exactly, although PLIER is known to be robust to initialization and to yield functionally equivalent decompositions across such runs~\cite{mao2019pathway}.

The main additional computational and memory overhead of FPLIER arises during initialization, where clients compute and securely aggregate the matrices $Y_iY_i^\top$. Initialization can become a bottleneck for very large $p$, since it scales as $\mathcal{O}(p^2)$, in contrast to centralized PLIER, which scales as $\mathcal{O}(pn)$. The dominant memory cost is also incurred at initialization: FPLIER stores the aggregated $YY^\top$ matrix with cost $\mathcal{O}(p^2)$, whereas centralized PLIER stores $Y$ with cost $\mathcal{O}(pn)$. Iterative costs are comparatively lower. A detailed complexity analysis is provided in Appendix~\ref{appendix:complexity}.

\subsubsection{Phase 3: postrelease membership inference risk assessment} 

After training, each client may optionally evaluate the sensitivity of the learned PLIER model by simulating a membership inference attack on its own data, as described in Section~\ref{sec:postrelease}. The attack follows the same intuition as the PCA-based analysis of Zari et al.~\cite{zari2022pca}: samples used during training are expected to be represented slightly better by the learned latent subspace than unseen samples.

If a client finds that the model is overly sensitive to its own data distribution, it may notify the server, block the model's release, and withdraw from the consortium.

\subsection{Membership inference attacks} 

We evaluate privacy risk through membership inference attacks, which test whether a specific sample was included in the training set. This attack class is particularly relevant in our setting because, unlike deep learning models, PLIER does not expose gradients or model updates that could enable gradient-based reconstruction or model inversion attacks. Instead, the information revealed during training consists of aggregated covariance-like statistics, from which membership information may still leak~\cite{zari2022pca}.

We study two attack settings: \emph{training-time} attacks, which exploit intermediate quantities observed during training and can be carried out only by the server, assumed to be \emph{honest-but-curious} (i.e., it follows the protocol correctly but may attempt to infer additional information from the data it observes); and \emph{postrelease} attacks, where an adversary leverages the released model parameters.

\begin{table*}[t]
\small
\centering
\caption{Training sample size, numerical rank of $Y$, and $n/p$ ratio across public data usage levels.}
\label{tab:ranks_all_methods}
\begin{tabular}{lccccccccc}
\toprule
& \multicolumn{3}{c}{\textbf{K-CLIER ($p=18{,}113$)}} & \multicolumn{3}{c}{\textbf{K-CLIER ($p=2{,}000$)}} & \multicolumn{3}{c}{\textbf{MultiPLIER ($p=6{,}574$)}} \\
\cmidrule(r){2-4} \cmidrule(r){5-7} \cmidrule(l){8-10}
\textbf{Public data used} & \textbf{$n$} & \textbf{$\operatorname{rank}(Y)$} & \textbf{$n/p$} & \textbf{$n$} & \textbf{$\operatorname{rank}(Y)$} & \textbf{$n/p$} & \textbf{$n$} & \textbf{$\operatorname{rank}(Y)$} & \textbf{$n/p$} \\
\midrule
0\%   & 927   & 927   & 0.051 & 927   & 927   & 0.464 & 937   & 937   & 0.143 \\
1\%   & 1,020 & 1,020 & 0.056 & 1,020 & 1,020 & 0.510 & 1,261 & 1,261 & 0.192 \\
5\%   & 1,392 & 1,392 & 0.077 & 1,392 & 1,392 & 0.696 & 2,556 & 2,553 & 0.389 \\
10\%  & 1,857 & 1,857 & 0.103 & 1,857 & 1,857 & 0.928 & 4,176 & 4,171 & 0.635 \\
15\%  & 2,322 & 2,322 & 0.128 & 2,322 & 2,000 & \textbf{1.161} & 5,796 & 5,787 & 0.882 \\
20\%  & 2,784 & 2,784 & 0.154 & 2,784 & 2,000 & \textbf{1.392} & 7,416 & 6,574 & \textbf{1.128} \\
50\%  & 5,570 & 5,569 & 0.308 & 5,570 & 2,000 & \textbf{2.785} & 17,132 & 6,574 & \textbf{2.606} \\
100\% & 10,211 & 10,204 & 0.564 & 10,211 & 2,000 & \textbf{5.106} & 33,327 & 6,574 & \textbf{5.070} \\
\bottomrule
\end{tabular}
\end{table*}

\subsubsection{Training-time membership inference attack}

During federated training, the honest-but-curious server observes only aggregated intermediate quantities produced via SecAgg+, namely the three matrix products in Eqs.~\eqref{eq:yyt} and \eqref{eq:bbtybt}, with the latter two recomputed at each iteration. Among these quantities, the covariance-like matrix $Y Y^\top$ poses the greatest privacy risk because it most directly reflects structure in the underlying gene expression profiles. By contrast, $B B^\top$ depends only on latent variables and is lower-dimensional by construction, while $Y B^\top$ mixes gene-level information with latent representations, making it harder to exploit for direct membership inference. We therefore focus on attacks targeting $Y Y^\top$. 

Prior work by Zari et al.~\cite{zari2022pca} shows that principal-component information can leak membership signals. In our setting, the server observes the full covariance-like matrix $Y Y^\top$, whose eigendecomposition reveals all principal component directions and associated variances for the training data $Y$ (Appendix~\ref{appendix:privacy}). This corresponds to the maximal-information regime considered in~\cite{zari2022pca}.

The attack proceeds as follows. Let $Y Y^\top = V \Lambda V^\top$ denote the eigendecomposition of the covariance-like matrix, where $\Lambda$ contains the nonnegative eigenvalues. Let $V_k \in \mathbb{R}^{p \times k}$ contain the top-$k$ eigenvectors. The attacker sets $k = \operatorname{rank}(YY^\top)$, thereby selecting all numerically nonzero principal components. It then forms the projector $\Pi_k = V_k V_k^\top$, projects a candidate sample as $\hat y = \Pi_k y$, and uses the reconstruction residual
\[
s_{\mathrm{train}}(y) = \|y - \hat y\|_2
\]
as a membership score. Intuitively, samples used during training tend to be represented slightly better by the principal-component subspace, yielding smaller residuals than non-members.

In our evaluation, we quantify attack effectiveness using the AUROC obtained from $s_{\mathrm{train}}(y)$. For each client $i$, we formulate a binary classification task in which the attacker attempts to distinguish \emph{members} from \emph{non-members}, treating the client’s private training samples $Y_i$ as members and its private holdout samples $Y_i^{\text{holdout}}$ as non-members.

\subsubsection{Postrelease membership inference attack} \label{sec:postrelease}

After training, an adversary with access to the released model may attempt to infer whether a specific sample was included in the training set. In our setting, the released objects include the gene loading matrix $Z$ and the associated ridge parameter $\lambda_2$, which together define the latent representation.

Given a candidate profile $y\in\mathbb{R}^p$, the adversary computes its latent representation via the ridge projection
\[
b(y) = (Z^\top Z + \lambda_2 I)^{-1} Z^\top y,
\]
reconstructs $\hat{y} = Z b(y)$, and uses the Spearman rank correlation 
\[
s_{\text{release}}(y) = \rho_{\mathrm{S}}(y,\hat{y})
\]
between $y$ and $\hat{y}$ as a membership score. We summarize membership-inference risk using the AUROC obtained by applying these scores to distinguish members from non-members, following the same labeling scheme as in the training-time attack.

In our client-side evaluation (Algorithm~\ref{alg:realworld_protocol}, Phase~3), we compute $s_{\text{release}}(y_{ij})$ for each private training sample and private holdout sample, and use the corresponding membership labels to estimate the AUROC locally for each client. A similar attack could also be constructed using a residual-based score, analogous to the training-time attack, by measuring the reconstruction error $\|y-\hat y\|_2$. In this work, we report results based on the Spearman correlation score because it aligns with the reconstruction-quality metrics previously used in the PLIER literature~\cite{malpetti2025protocol}. Nevertheless, we verified that using the residual norm yields qualitatively similar results.

\section{Experiments}
\label{sec:experiments}

FPLIER is designed to preserve the algebraic structure of the centralized PLIER updates (Appendix~\ref{appendix:training}). Accordingly, our experimental evaluation has two main objectives. First, we assess whether this update-level equivalence yields decompositions of comparable quality to those of centralized PLIER under realistic federated conditions, including heterogeneous client partitions and secure aggregation. Second, we evaluate the robustness of the resulting models to membership-inference attacks, both during training and after model release, with particular emphasis on how privacy risk varies with the amount of public data incorporated during training.

\subsection{Datasets}

To evaluate FPLIER, we used resources from the K--CLIER~\cite{clier} and MultiPLIER~\cite{taroni2019multiplier} studies. In both settings, we relied on the original materials provided with the studies, including the gene set and gene expression matrices (in RPKM units).

In the K--CLIER setting, the gene set matrix comprises 726 gene sets derived from kidney-related single-cell transcriptomic signatures. Gene expression data were obtained from The Cancer Genome Atlas (TCGA)~\cite{tcga}, spanning 33 cancer types and totaling 11{,}348 samples. In the MultiPLIER setting, the gene set matrix contains 628 Reactome-derived gene sets~\cite{fabregat2018reactome}. Gene expression data were drawn from the Recount2 compendium~\cite{collado2017reproducible}, which contains 37{,}032 uniformly processed human RNA-seq samples from 1{,}466 studies, with study sizes ranging from 1 to more than 1{,}700 samples.

\subsection{Experimental design}

To evaluate FPLIER, we constructed simulated consortia using the datasets described  (Table~\ref{tab:cohorts}). The two settings are intentionally complementary: K--CLIER represents a focused, domain-specific use case with a specialized prior and a more limited public compendium, whereas MultiPLIER represents a broader, more heterogeneous scenario with substantially larger public-data support. Together, they provide a controlled and interpretable environment that captures a realistic use case for FPLIER: combining a large public transcriptomic compendium with smaller, domain-specific cohorts that remain locally hosted by different clients.

We conducted a series of federated simulations in which the proportion of available public data used during training was varied at levels of 0\%, 1\%, 5\%, 10\%, 15\%, 20\%, 50\%, and 100\%. For each configuration, the full protocol in Algorithm~\ref{alg:realworld_protocol} was executed independently, using a 90/10 split of each client’s local dataset into training and holdout subsets. In the K--CLIER setting, we additionally considered a reduced-dimensionality configuration restricted to the 2{,}000 most variable genes. In parallel, for each configuration, a centralized PLIER model was trained on the pooled training data, serving as a ground-truth reference for benchmarking the federated models.

Using these models, we conducted two analyses. First, we assessed whether FPLIER reproduces the behavior of centralized PLIER in heterogeneous federated settings by analyzing convergence and decomposition quality. Second, we evaluated membership-inference risk under different training configurations, varying both the amount of public data incorporated during training and, in the K--CLIER setting, the reduced 2{,}000-gene configuration. We considered both training-time attacks from the server's perspective and postrelease attacks conducted by an external adversary.

Feature dimensionalities, training sample sizes, empirical ranks, and the sample-to-dimension ratios ($n/p$) for the two K--CLIER settings and the MultiPLIER setting at each public-data level are reported in Table~\ref{tab:ranks_all_methods}. In high-dimensional gene expression data, samples can generally be expected to be approximately linearly independent, implying that the rank of the expression matrix is close to $\min(n,p)$. The empirical ranks in Table~\ref{tab:ranks_all_methods} are consistent with this expectation, indicating that $n/p$ provides a practical proxy for proximity to the full-rank regime while being considerably easier to compute than the global matrix rank in a federated setting. In particular, values of $n/p < 1$ correspond to a rank-deficient regime, $n/p \approx 1$ to the transition toward full rank, and $n/p > 1$ to a full-rank regime. In the following, we therefore report membership-inference performance as a function of $n/p$.

\subsection{Effectiveness of FPLIER decomposition}

Across all experimental settings, both centralized PLIER and FPLIER converged reliably. To compare the quality of the resulting models, we assessed reconstruction fidelity~\cite{malpetti2025protocol}. At convergence, we evaluated the reconstructed expression matrix $\hat{Y} = ZB$ for each model and, for every gene, computed the Spearman correlation between its reconstructed and original expression profile, yielding a distribution of gene-wise correlation coefficients for each decomposition. A Kolmogorov–Smirnov test comparing the centralized and federated distributions across all configurations did not detect statistically significant differences, indicating that reconstruction performance is statistically indistinguishable between PLIER and FPLIER.

\subsection{Membership inference attacks}

\begin{figure*}[t]
    \centering    \includegraphics[width=0.99\textwidth]{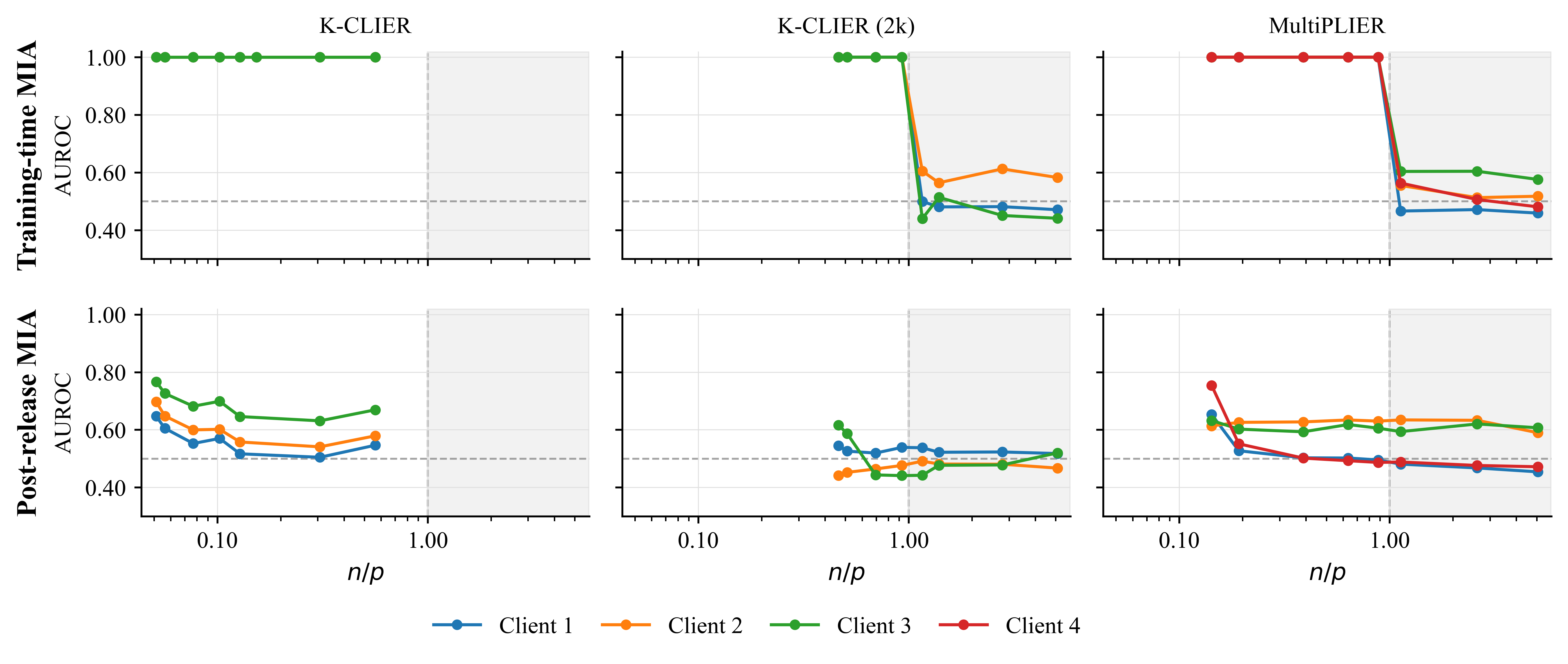}
    \caption{Membership inference attack performance during training (left) and after model release (right) as a function of the sample-to-feature dimensionality ($n/p$) ratio for the K--CLIER, K--CLIER (2k genes), and MultiPLIER settings. The dashed horizontal line indicates random guessing (AUROC $=0.5$). As $n/p$ increases, attack performance decreases toward random guessing, indicating reduced privacy risk in higher-rank regimes for both attacks. In the top-left panel, the curves are nearly perfectly superimposed and thus visually indistinguishable.}    \label{fig:mia_two_panel}
\end{figure*}

We evaluated the training-time and postrelease membership inference attacks described in Section~\ref{sec:methods}. Figure~\ref{fig:mia_two_panel} reports the resulting AUROC values as a function of the $n/p$ ratio across the K--CLIER, K--CLIER (2k genes), and MultiPLIER settings. Across all settings, the strongest attacks occur when $n/p$ is small, corresponding to configurations in which little or no public data are incorporated during training. In this rank-deficient regime, both the covariance-like statistics exposed during training and the released model retain substantial membership signal. The training-time attack achieves AUROC values close to one, while the postrelease attack remains well above random guessing.

For the training-time attack, the dominant empirical pattern is a sharp drop in AUROC as the system approaches the full-rank regime ($n/p \approx 1$). This transition is particularly evident in the K--CLIER (2k genes) and MultiPLIER settings, where attack performance is near-perfect for $n/p < 1$ and then rapidly collapses toward random guessing (AUROC $\approx 0.5$) once $n/p$ exceeds approximately one. In the original K--CLIER setting, all configurations remain in the rank-deficient regime, and the attack is always highly effective. 

The postrelease attack exhibits the same overall dependence on $n/p$, but with notable differences. AUROC values in the rank-deficient regime are generally lower, the decline as additional samples are incorporated is more gradual, and the results are noisier across clients, so the trajectories are not always strictly monotone decreasing. A notable pattern in the MultiPLIER setting is that two clients exhibit nearly flat AUROC trajectories as $n/p$ increases. For one of them, the holdout set contains roughly 40 samples, making finite-sample variability an unlikely explanation. Repeating the experiment with a different train/holdout split seed reproduces the same pattern, suggesting that it may instead reflect distributional or other data-specific effects.

Reducing the dimensionality, as in the K--CLIER (2k genes) setting, shifts the system into a higher-$n/p$ regime at substantially smaller sample sizes. As a result, the transition to the full-rank regime occurs earlier, and both attacks become markedly less effective than in the original K--CLIER configuration.

These results indicate that membership-inference risk is governed primarily by proximity to the full-rank regime of the training expression data, and that increasing public-data support or reducing dimensionality substantially mitigates attack effectiveness.

\section{Discussion}\label{discussion}

A central question in applying federated learning to bulk transcriptomic data is whether their sensitivity truly justifies the added complexity of FL, or whether such methods risk being disproportionate for data often regarded as relatively low-risk. Although prior work in federated transcriptomics typically treats privacy and governance constraints as a given motivation, the actual sensitivity of bulk expression profiles and the corresponding threat model are often examined only partially. Clarifying this issue is important for interpreting both the design choices underlying FPLIER and the empirical results reported here. We therefore first discuss the privacy and governance considerations specific to bulk transcriptomic data, then situate FPLIER within this broader context, and finally outline the main limitations of the present study.

\subsection{Privacy and governance considerations for bulk transcriptomics}

Compared to other biomedical data types, such as whole-genome 
sequencing or raw genotype data, bulk RNA-seq expression profiles are generally considered less sensitive. They do not encode stable, uniquely identifying variants, and reflect contextual factors (e.g., tissue, disease state, experimental conditions) that may vary over short periods. This makes them impractical to cross-reference with other clinical or genetic information, even when such information is available to an attacker. As a result, transcriptomic data are often viewed as carrying a lower intrinsic risk of re-identification~\cite{Zhou_2024}.

Nonetheless, transcriptomic profiles cannot be considered risk-free. Prior work on biomedical data privacy has shown that high-dimensional molecular measurements can still enable inference attacks under certain conditions, particularly when expression profiles can be linked to auxiliary information. For instance, they can be used to infer single-nucleotide variants, which in turn can determine with high certainty whether individuals with known sequence information were members of a gene expression study cohort~\cite{schadt}. Thus, relevant risks include membership inference, attribute inference, and linkage attacks, whose effectiveness depends strongly on dimensionality, sample size, and the availability of external reference datasets~\cite{Zhou_2024}. While these risks are not as pronounced as for genetic and genomic data, they remain non-negligible.

The practical sensitivity of transcriptomic data also depends on regulatory and organizational context. Even when the technical re-identification risk is limited, governance frameworks may restrict cross-institutional data sharing, especially in clinical and translational research environments. As noted in~\cite{Slobogin_2025}, existing legal protections for biomedical data are fragmented and include numerous exceptions, prompting many institutions to adopt conservative data-sharing policies even for data types that are not inherently identifying. In these situations, the inability to centralize transcriptomic datasets stems as much from institutional and legal constraints as from technical privacy concerns.

From this perspective, federated learning offers a pragmatic and proportionate solution. Rather than assuming transcriptomic data are maximally sensitive, FL enables collaboration under a conservative threat model that limits exposure to both technical inference attacks and governance-related risks, while still allowing institutions to benefit from large and diverse training cohorts. Its design is naturally aligned with regulatory principles, such as those embedded in the GDPR and the AI Act, and may therefore be preferred by institutions adopting a cautious approach to data sharing~\cite{malpetti2025technical}.

\subsection{Privacy considerations for FPLIER} \label{sec:privacy_discussion}

FPLIER is designed under an honest-but-curious threat model, in which the server and the clients follow the protocol correctly but may attempt to infer additional information from the quantities they are allowed to observe. In this setting, privacy risk arises at two distinct stages: during training, where an honest-but-curious server may attempt to infer membership from aggregated covariance-like statistics, and after training, where an external adversary may attempt to infer membership using the released model.

The clearest empirical pattern in our study is the sharp reduction in training-time membership-inference performance as the system transitions from a rank-deficient to a full-rank regime. This behavior is consistent with the mechanics of PCA-based attacks. When the number of samples is small relative to the feature dimension, the data matrix is necessarily rank-deficient, and the training samples lie in a lower-dimensional subspace that can be recovered from the principal components of the covariance-like matrix. Because this subspace is estimated directly from the training data, training samples tend to be represented unusually well by it, yielding systematically smaller projection residuals than those of non-members, thereby enabling highly effective membership inference.

As the number of samples increases and the training matrix approaches the full-rank regime, no low-dimensional subspace can describe the training samples well enough to fully identify them. Indeed, samples drawn from the same underlying distribution tend to produce similar projection residuals whether or not they were included in training. This provides a principled explanation for the sharp drop in AUROC observed for the training-time attack once $n/p$ approaches or exceeds one. Reaching full rank, however, does not imply that attack performance becomes exactly equal to random guessing, since it still depends on the data distribution and finite-sample effects, although the observed AUROC values approach this level.

These results suggest a practical deployment guideline: consortia should aim to operate in a regime in which the aggregated training matrix is effectively full rank. Because estimating the global rank directly may not be straightforward in a federated setting, the ratio $n/p$ serves as a simple, useful operational proxy. In practice, a consortium can increase $n/p$ either by incorporating additional public data or by reducing feature dimensionality before training, aiming to achieve values of $n/p$ well above 1.

A further protective aspect of the protocol is that, although the server knows which public samples are assigned to each client, it does not know which subset is used for local training, since each client’s train/holdout split remains private. This design reduces the effectiveness of server-side attacks based on $YY^\top$. If the server knew exactly which public samples had been incorporated into training, it could, in principle, subtract their contribution from $YY^\top$ and attack the reduced matrix $YY^\top - Y^{\mathrm{pub}} (Y^{\mathrm{pub}})^\top$, 
which would typically be of a lower rank and therefore more vulnerable to membership inference. Because the server does not know which public samples were retained in each local training split, such subtraction cannot be performed exactly in practice. Moreover, once the system is in a full-rank regime, the residuals themselves provide little information for reconstructing the hidden allocation. This helps preserve the protective effect of public support and provides an additional layer of defense against server-side inference.

The same intuition suggests some robustness beyond the honest-but-curious setting. If some clients collude with the server, their contributions can be subtracted from $YY^\top$. 
In some cases, the resulting matrix may become rank-deficient, making it vulnerable to membership inference; if it remains full-rank, however, it reveals little useful membership information about the non-colluding parties. In the limiting case where all local gene expression matrices are full-rank, the attack remains ineffective for any client, regardless of the number of colluding clients. Therefore, reducing feature dimensionality or increasing public-data support can improve robustness (in our experiments, the 100\% public-data usage scenarios for K--CLIER 2k and MultiPLIER are robust to collusion by an arbitrary number of clients). However, in realistic usage scenarios, such measures may not always be sufficient, especially in consortia with a large number of clients.

In the postrelease setting, the same broad trend is observed as for the training-time attack: membership-inference performance generally decreases as the rank of the training data increases. The interpretation is less direct, however, because the released transformation is no longer an orthogonal PCA projector. Instead, the learned representation is shaped by both regularization and prior biological knowledge through the gene-set constraints. As a result, reconstruction quality need not be uniform across the data space. For some clients or biological contexts, the learned latent space may align more closely with the general structure of their samples, yielding a more refined representation in those regions. In such cases, the released model may still distinguish between training and holdout samples to some extent for those clients, even when the overall system is full-rank, and the $n/p$ ratio exceeds 1. This may help explain the more heterogeneous postrelease behavior observed across clients, including the nearly flat AUROC trajectories in the MultiPLIER setting, although further analysis would be needed to confirm this interpretation.

The postrelease attack is more concerning than the training-time attack because it can be carried out by any party with access to the released model. However, FPLIER provides a practical safeguard by enabling an empirical assessment of membership-inference risk prior to release. Each client evaluates the model locally on its own training and holdout samples and can detect whether the learned representation captures signals that are overly specific to its data distribution. In the regimes identified above, where the training matrix is full-rank, any such residual sensitivity is expected to be small. Nevertheless, if a client observes unexpectedly high sensitivity, this information remains within the consortium, and the client can request the model release to be blocked before any external disclosure occurs. The consortium may then revise the training configuration by incorporating additional public data or reducing feature dimensionality, or decide not to release the model at all. 

\subsection{Limitations}

This study has several limitations. First, FPLIER is analyzed under an honest-but-curious threat model, in which the server and clients follow the protocol correctly but may attempt to infer additional information from the quantities they observe. Stronger adversarial settings, including malicious protocol deviations, collusion between parties, and active poisoning attacks, are outside the scope of the present work. Second, the privacy protection provided here is empirical rather than formal, as would be the case, for example, when using differential privacy. Finally, our evaluation is based on simulated federated consortia in two transcriptomic settings and on scenarios in which relevant public data are available; although these can be considered prototypical, further validation in additional deployments would help confirm the generality of the observed privacy–utility trade-offs.

\section{Conclusions and future work} \label{sec:conclusions}

We introduced FPLIER, a federated reformulation of PLIER that enables collaborative training across multiple data holders while keeping gene expression data local. Across two simulated transcriptomic consortia, FPLIER converged reliably and produced decompositions whose quality was statistically indistinguishable from those obtained with centralized PLIER. We further showed that membership-inference risk is governed primarily by the rank regime of the training expression matrix and can be substantially reduced by incorporating public data or reducing dimensionality.

Several extensions remain for future work. One direction is to further reduce the exposure of sensitive covariance-like statistics during initialization, for example, by leveraging federated singular value decomposition methods. Several such approaches have been proposed in the literature~\cite{hartebrodt2022federated}, although they are not currently available in the Flower framework used in this work. A complementary direction is to incorporate differential privacy mechanisms, which could provide formal privacy guarantees and strengthen the overall robustness of the framework.

\section{Code availability}
\label{sec:code_availability}

The implementation of FPLIER, together with documentation and a usage example, is available at \href{https://github.com/IDSIA/FPLIER}{https://github.com/IDSIA/FPLIER}.

\anontext{}{\section{Acknowledgments}
The authors thank Emanuele Delucchi for insightful discussions on the mathematical aspects of membership inference attacks. This work was partially supported by the European Union Horizon 2020 programme [101136962]; UK Research and Innovation (UKRI) under the UK Government’s Horizon Europe funding guarantee [10098097, 10104323] and the Swiss State Secretariat for Education, Research and Innovation (SERI).}

\printbibliography

@String{Computing = "Computing" }

@String{Computer = "{IEEE} Computer" }

@String{Springer = "Springer-Verlag" }

@article{barbie2009systematic,
  title={Systematic RNA interference reveals that oncogenic KRAS-driven cancers require TBK1},
  author={Barbie, David A and Tamayo, Pablo and Boehm, Jesse S and Kim, So Young and Moody, Susan E and Dunn, Ian F and Schinzel, Anna C and Sandy, Peter and Meylan, Etienne and Scholl, Claudia and others},
  journal={Nature},
  volume={462},
  number={7269},
  pages={108--112},
  year={2009},
  publisher={Nature Publishing Group UK London}
}

@ArtifactSoftware{R,
    title = {R: A Language and Environment for Statistical Computing},
    author = {{R Core Team}},
    organization = {R Foundation for Statistical Computing},
    address = {Vienna, Austria},
    year = {2019},
    url = {https://www.R-project.org/},
}

@article{mcmahan2016communication,
  title={Communication-efficient learning of deep networks from decentralized data. arXiv},
  author={McMahan, H Brendan and Moore, Eider and Ramage, Daniel and Hampson, Seth and Arcas, BA},
  journal={arXiv preprint arXiv:1602.05629},
  year={2016}
}

@ARTICLE{mao2019pathway,
  title={Pathway-level information extractor (PLIER) for gene expression data},
  author={Mao, Weiguang and Zaslavsky, Elena and Hartmann, Boris M and Sealfon, Stuart C and Chikina, Maria},
  journal={Nature methods},
  volume={16},
  number={7},
  pages={607--610},
  year={2019},
  publisher={Nature Publishing Group US New York},
  %doi={10.1038/s41592-019-0456-1}
}

@ARTICLE{taroni2019multiplier,
  title={MultiPLIER: a transfer learning framework for transcriptomics reveals systemic features of rare disease},
  author={Taroni, Jaclyn N and Grayson, Peter C and Hu, Qiwen and Eddy, Sean and Kretzler, Matthias and Merkel, Peter A and Greene, Casey S},
  journal={Cell systems},
  volume={8},
  number={5},
  pages={380--394},
  year={2019},
  publisher={Elsevier}
}

@article{wang2024scfed,
  title={scFed: federated learning for cell type classification with scRNA-seq},
  author={Wang, Shuang and Shen, Bochen and Guo, Lanting and Shang, Mengqi and Liu, Jinze and Sun, Qi and Shen, Bairong},
  journal={Briefings in Bioinformatics},
  volume={25},
  number={1},
  pages={bbad507},
  year={2024},
  publisher={Oxford University Press}
}

@article{sav2022privacy,
  title={Privacy-preserving federated neural network learning for disease-associated cell classification},
  author={Sav, Sinem and Bossuat, Jean-Philippe and Troncoso-Pastoriza, Juan R and Claassen, Manfred and Hubaux, Jean-Pierre},
  journal={Patterns},
  volume={3},
  number={5},
  year={2022},
  publisher={Elsevier}
}

@article{clier,
  title={A transfer learning framework to elucidate the clinical relevance of altered proximal tubule cell states in kidney disease},
  author={Legouis, David and Rinaldi, Anna and Malpetti, Daniele and Arnoux, Gregoire and Verissimo, Thomas and Faivre, Anna and Mangili, Francesca and Rinaldi, Andrea and Ruinelli, Lorenzo and Pugin, Jerome and others},
  journal={Iscience},
  volume={27},
  number={3},
  year={2024},
  publisher={Elsevier}
}

@article{halko2011finding,
  title={Finding structure with randomness: Probabilistic algorithms for constructing approximate matrix decompositions},
  author={Halko, Nathan and Martinsson, Per-Gunnar and Tropp, Joel A},
  journal={SIAM review},
  volume={53},
  number={2},
  pages={217--288},
  year={2011},
  publisher={SIAM}
}

@article{hanzelmann2013gsva,
  title={GSVA: gene set variation analysis for microarray and RNA-seq data},
  author={H{\"a}nzelmann, Sonja and Castelo, Robert and Guinney, Justin},
  journal={BMC bioinformatics},
  volume={14},
  pages={1--15},
  year={2013},
  publisher={Springer}
}

@article{collado2017reproducible,
  title={Reproducible RNA-seq analysis using recount2},
  author={Collado-Torres, Leonardo and Nellore, Abhinav and Kammers, Kai and Ellis, Shannon E and Taub, Margaret A and Hansen, Kasper D and Jaffe, Andrew E and Langmead, Ben and Leek, Jeffrey T},
  journal={Nature biotechnology},
  volume={35},
  number={4},
  pages={319--321},
  year={2017},
  publisher={Nature Publishing Group US New York}
}

@article{zhang2024mousiplier,
  title={MousiPLIER: A Mouse Pathway-Level Information Extractor Model},
  author={Zhang, Shuo and Heil, Benjamin J and Mao, Weiguang and Chikina, Maria and Greene, Casey S and Heller, Elizabeth A},
  journal={eneuro},
  volume={11},
  number={6},
  year={2024},
  publisher={Society for Neuroscience}
}

@article{fabregat2018reactome,
  title={The reactome pathway knowledgebase},
  author={Fabregat, Antonio and Jupe, Steven and Matthews, Lisa and Sidiropoulos, Konstantinos and Gillespie, Marc and Garapati, Phani and Haw, Robin and Jassal, Bijay and Korninger, Florian and May, Bruce and others},
  journal={Nucleic acids research},
  volume={46},
  number={D1},
  pages={D649--D655},
  year={2018},
  publisher={Oxford University Press}
}

@inproceedings{bell,
  title={Secure single-server aggregation with (poly) logarithmic overhead},
  author={Bell, James Henry and Bonawitz, Kallista A and Gasc{\'o}n, Adri{\`a} and Lepoint, Tancr{\`e}de and Raykova, Mariana},
  booktitle={Proceedings of the 2020 ACM SIGSAC conference on computer and communications security},
  pages={1253--1269},
  year={2020}
}

@inproceedings{li21,
  title={Secure aggregation for federated learning in flower},
  author={Li, Kwing Hei and de Gusm{\~a}o, Pedro Porto Buarque and Beutel, Daniel J and Lane, Nicholas D},
  booktitle={Proceedings of the 2nd ACM International Workshop on Distributed Machine Learning},
  pages={8--14},
  year={2021}
}

@inproceedings{zari2022pca,
  title={Membership inference attack against principal component analysis},
  author={Zari, Oualid and Parra-Arnau, Javier and {\"U}nsal, Ay{\c{s}}e and Strufe, Thorsten and {\"O}nen, Melek},
  booktitle={International Conference on Privacy in Statistical Databases},
  pages={269--282},
  year={2022},
  organization={Springer}
}

@article{malpetti2025protocol,
  title={Protocol for interpretable and context-specific single-cell-informed deconvolution of bulk RNA-seq data},
  author={Malpetti, Daniele and Mangili, Francesca and Bolis, Marco and Rinaldi, Anna and Legouis, David and Ruinelli, Lorenzo and Cipp{\`a}, Pietro and Azzimonti, Laura},
  journal={STAR protocols},
  volume={6},
  number={1},
  pages={103670},
  year={2025},
  publisher={Elsevier}
}

@article{malpetti2025technical,
  title={Technical and legal aspects of federated learning in bioinformatics: applications, challenges and opportunities},
  author={Malpetti, Daniele and Scutari, Marco and Gualdi, Francesco and Van Setten, Jessica and van der Laan, Sander W and Haitjema, Saskia and Lee, Aaron and Hering, Isabelle and Mangili, Francesca},
  journal={Frontiers in Digital Health},
  volume={7},
  pages={1644291},
  year={2025},
  publisher={Frontiers}
}

@article{Zhou_2024,
  title={Patient privacy in AI-driven omics methods},
  author={Zhou, Juexiao and Huang, Chao and Gao, Xin},
  journal={Trends in Genetics},
  volume={40},
  number={5},
  pages={383--386},
  year={2024},
  publisher={Elsevier}
}

@article{Slobogin_2025,
  title={A decade of research on genetic privacy: the findings of the GetPreCiSe Center at Vanderbilt University},
  author={Slobogin, Christopher and Tellis, Karli and Clayton, Ellen Wright and Clayton, Jay and Eilmus, Ayden and Malin, Bradley A},
  journal={Frontiers in Genetics},
  volume={16},
  pages={1629386},
  year={2025},
  publisher={Frontiers Media SA}
}

@article{flower,
  title={Flower: A friendly federated learning research framework},
  author={Beutel, Daniel J and Topal, Taner and Mathur, Akhil and Qiu, Xinchi and Fernandez-Marques, Javier and Gao, Yan and Sani, Lorenzo and Li, Kwing Hei and Parcollet, Titouan and de Gusm{\~A}{\c{G}}o, Pedro Porto Buarque and others},
  journal={arXiv preprint arXiv:2007.14390},
  year={2020}
}

@article{riedel,
  title={Comparative analysis of open-source federated learning frameworks-a literature-based survey and review},
  author={Riedel, Pascal and Schick, Lukas and von Schwerin, Reinhold and Reichert, Manfred and Schaudt, Daniel and Hafner, Alexander},
  journal={International Journal of Machine Learning and Cybernetics},
  volume={15},
  number={11},
  pages={5257--5278},
  year={2024},
  publisher={Springer}
}

@article{wang2009rna,
  title={RNA-Seq: a revolutionary tool for transcriptomics},
  author={Wang, Zhong and Gerstein, Mark and Snyder, Michael},
  journal={Nature reviews genetics},
  volume={10},
  number={1},
  pages={57--63},
  year={2009},
  publisher={Nature Publishing Group UK London}
}

@article{stubbington2017single,
  title={Single-cell transcriptomics to explore the immune system in health and disease},
  author={Stubbington, Michael JT and Rozenblatt-Rosen, Orit and Regev, Aviv and Teichmann, Sarah A},
  journal={Science},
  volume={358},
  number={6359},
  pages={58--63},
  year={2017},
  publisher={American Association for the Advancement of Science}
}

@article{clarke2008properties,
  title={The properties of high-dimensional data spaces: implications for exploring gene and protein expression data},
  author={Clarke, Robert and Ressom, Habtom W and Wang, Antai and Xuan, Jianhua and Liu, Minetta C and Gehan, Edmund A and Wang, Yue},
  journal={Nature reviews cancer},
  volume={8},
  number={1},
  pages={37--49},
  year={2008},
  publisher={Nature Publishing Group UK London}
}

@article{cai2025federated,
  title={Federated deep learning enables cancer subtyping by proteomics},
  author={Cai, Zhaoxiang and Boys, Emma L and Noor, Zainab and Aref, Adel T and Xavier, Dylan and Lucas, Natasha and Williams, Steven G and Koh, Jennifer MS and Poulos, Rebecca C and Wu, Yangxiu and others},
  journal={Cancer Discovery},
  volume={15},
  number={9},
  pages={1803--1818},
  year={2025},
  publisher={American Association for Cancer Research}
}

@article{li2022federated,
  title={Federated learning algorithms for generalized mixed-effects model (GLMM) on horizontally partitioned data from distributed sources},
  author={Li, Wentao and Tong, Jiayi and Anjum, Md Monowar and Mohammed, Noman and Chen, Yong and Jiang, Xiaoqian},
  journal={BMC Medical Informatics and Decision Making},
  volume={22},
  number={1},
  pages={269},
  year={2022},
  publisher={Springer}
}

@article{uffelmann2021genome,
  title={Genome-wide association studies},
  author={Uffelmann, Emil and Huang, Qin Qin and Munung, Nchangwi Syntia and De Vries, Jantina and Okada, Yukinori and Martin, Alicia R and Martin, Hilary C and Lappalainen, Tuuli and Posthuma, Danielle},
  journal={Nature Reviews Methods Primers},
  volume={1},
  number={1},
  pages={59},
  year={2021},
  publisher={Nature Publishing Group UK London}
}

@article{schadt,
  title={Bayesian method to predict individual SNP genotypes from gene expression data},
  author={Schadt, Eric E and Woo, Sangsoon and Hao, Ke},
  journal={Nature genetics},
  volume={44},
  number={5},
  pages={603--608},
  year={2012},
  publisher={Nature Publishing Group US New York}
}

@article{zolotareva2021flimma,
  title={Flimma: a federated and privacy-aware tool for differential gene expression analysis},
  author={Zolotareva, Olga and Nasirigerdeh, Reza and Matschinske, Julian and Torkzadehmahani, Reihaneh and Bakhtiari, Mohammad and Frisch, Tobias and Sp{\"a}th, Julian and Blumenthal, David B and Abbasinejad, Amir and Tieri, Paolo and others},
  journal={Genome biology},
  volume={22},
  number={1},
  pages={338},
  year={2021},
  publisher={Springer}
}

@article{dwork2014algorithmic,
  title={The algorithmic foundations of differential privacy},
  author={Dwork, Cynthia and Roth, Aaron},
  journal={Foundations and trends in theoretical computer science},
  volume={9},
  number={3-4},
  pages={211--487},
  year={2014},
  publisher={Emerald Publishing Limited}
}

@article{hartebrodt2022federated,
  title={Federated horizontally partitioned principal component analysis for biomedical applications},
  author={Hartebrodt, Anne and R{\"o}ttger, Richard},
  journal={Bioinformatics Advances},
  volume={2},
  number={1},
  pages={vbac026},
  year={2022},
  publisher={Oxford University Press}
}

@inproceedings{bonawitz2017practical,
  title={Practical secure aggregation for privacy-preserving machine learning},
  author={Bonawitz, Keith and Ivanov, Vladimir and Kreuter, Ben and Marcedone, Antonio and McMahan, H Brendan and Patel, Sarvar and Ramage, Daniel and Segal, Aaron and Seth, Karn},
  booktitle={proceedings of the 2017 ACM SIGSAC Conference on Computer and Communications Security},
  pages={1175--1191},
  year={2017}
}

@book{golub,
author = {Golub, Gene H. and Van Loan, Charles F.},
title = {Matrix computations (3rd ed.)},
year = {1996},
isbn = {0801854148},
publisher = {Johns Hopkins University Press},
address = {USA}
}

@article{tcga,
  title={The cancer genome atlas pan-cancer analysis project},
  author={Weinstein, John N and Collisson, Eric A and Mills, Gordon B and Shaw, Kenna R and Ozenberger, Brad A and Ellrott, Kyle and Shmulevich, Ilya and Sander, Chris and Stuart, Joshua M},
  journal={Nature genetics},
  volume={45},
  number={10},
  pages={1113--1120},
  year={2013},
  publisher={Nature Publishing Group}
}

\clearpage

\appendix

\section{Standard PLIER training} \label{appendix:plier}

For completeness, we briefly summarize the standard centralized training procedure of PLIER, following the original formulation by Mao et al.~\cite{mao2019pathway}. The purpose of this appendix is to provide a reference point for comparison with the federated variant introduced in Appendix~\ref{appendix:training} and to highlight which parts of the training procedure must be reformulated to extend to the distributed setting.

Algorithm~\ref{alg:plier} presents a simplified but faithful summary of the core PLIER optimization scheme. Starting from the expression matrix $Y$ and the gene-set matrix $C$, PLIER first uses a singular value decomposition (SVD) of $Y$ to determine the latent dimensionality $k$, determine the regularization parameters $\lambda_1$ and $\lambda_2$, and initialize the latent-representation matrix $B$. Training then proceeds through alternating updates of $Z$, $B$, and $U$. In this decomposition, $Z$ captures gene loadings, $B$ contains the latent representation of the samples, and $U$ links latent variables to prior gene sets.

The original implementation includes several additional practical refinements, such as a more specialized sparse update for $U$ and an optional adjustment of $\lambda_3$ during training. We omit these details here because they are not essential for understanding the algebraic structure of PLIER or the way that structure is adapted in the federated setting.

\section{Federated implementation details}
\label{appendix:training_appendix}

This appendix provides additional detail on the federated implementation of FPLIER. In particular, it clarifies how consortium-wide gene-wise standardization is performed, how the centralized PLIER optimization is distributed across clients and the server, which quantities remain local, and which must be obtained through secure aggregation.

For readability, the pseudocode explicitly indicates the actor responsible for each step, distinguishing among \textbf{Client}-, \textbf{Server}-, and \textbf{Consortium}-level operations. When a quantity is said to be computed ``via SecAgg+'', this means that clients submit local contributions through secure aggregation, and the server observes only the resulting aggregate rather than the individual client contributions.

\subsection{Federated standardization}
\label{appendix:standardization}

\begin{algorithm}[b]
\small
\caption{Standard PLIER training}
\label{alg:plier}
\begin{algorithmic}[1]
\Require Gene expression matrix $Y \in \mathbb{R}^{p \times n}$; gene set matrix $C$; maximum number of iterations $T$; tolerance $\tau$
\Ensure PLIER model $(C,Z,U,\lambda_2)$

\Statex \hspace{-\algorithmicindent}\textit{Initialization}
\State Compute randomized SVD $Y \simeq V D W^\top$
\State Select $k$ and $\lambda_2$ via an elbow heuristic on $D$
\State Set $\lambda_1 \gets \lambda_2/2$
\State Initialize $\lambda_3$ to a small positive value
\State Initialize $B^{(0)} \gets (W_{[:,1:k]} D_{1:k})^\top$
\State Initialize $U^{(0)} \gets 0$

\Statex \hspace{-\algorithmicindent}\textit{Alternating updates}
\For{$\ell = 0,1,\dots,T-1$}

  \Statex \hspace{\algorithmicindent}\textit{$Z$ update}
  \State $Z^{(\ell+1)} \gets
  \big(Y {B^{(\ell)}}^\top + \lambda_1 C U^{(\ell)}\big)
  \big(B^{(\ell)} {B^{(\ell)}}^\top + \lambda_1 I\big)^{-1}$

  \State $Z^{(\ell+1)}[Z^{(\ell+1)}<0] \gets 0$

  \Statex \hspace{\algorithmicindent}\textit{$B$ update}

  \State $B^{(\ell+1)} \gets
  \big({Z^{(\ell+1)}}^\top Z^{(\ell+1)} + \lambda_2 I\big)^{-1}
  {Z^{(\ell+1)}}^\top Y$

    \Statex \hspace{\algorithmicindent}\textit{$U$ update}

\State $U^{(\ell+1)} \gets
  \arg\min_U \|Z^{(\ell+1)} - C U\|_F^2 + \lambda_3 \|U\|_1$

  \State Update $\lambda_3$ to control  $U^{(\ell+1)}$ sparsity (optional)

    \Statex \hspace{\algorithmicindent}\textit{Early stop}

    \State Stop if $\|B^{(\ell+1)} - B^{(\ell)}\|_F^2 / \|B^{(\ell+1)}\|_F^2 < \tau$

\EndFor
\end{algorithmic}
\end{algorithm}

\begin{algorithm}[b]
\small
\caption{Federated gene-wise standardization}
\label{alg:fed_standardization}
\begin{algorithmic}[1]
\Require Client training matrices $\{Y_i\}_{i=1}^N$
\Ensure Standardized matrices $\{Y_i\}_{i=1}^N$

\ForAll{genes $g \in \{1,\dots,p\}$ \textbf{in parallel}}

\Statex \hspace{-\algorithmicindent}\textit{Local statistics}
\ForAll{$i \in \{1,\dots,N\}$ \textbf{in parallel}}
  \State \textbf{Client $i$:} Compute $n_i \gets$ number of samples in $Y_i$
  \State \textbf{Client $i$:} Compute $s_i \gets \sum_j Y_{i,gj}$
  \State \textbf{Client $i$:} Compute $q_i \gets \sum_j Y_{i,gj}^2$
\EndFor

\Statex \hspace{-\algorithmicindent}\textit{Secure aggregation}
\State \textbf{Consortium:} Compute $n \gets \sum_{i=1}^N n_i$ via SecAgg+
\State \textbf{Consortium:} Compute $s \gets \sum_{i=1}^N s_i$ via SecAgg+
\State \textbf{Consortium:} Compute $q \gets \sum_{i=1}^N q_i$ via SecAgg+

\Statex \hspace{-\algorithmicindent}\textit{Global statistics}
\State \textbf{Server:} Compute mean $\mu_g \gets s/n$
\State \textbf{Server:} Compute standard deviation $\sigma_g \gets \sqrt{q/n - \mu_g^2}$
\State \textbf{Server:} Broadcast $(\mu_g,\sigma_g)$ to all clients

\Statex \hspace{-\algorithmicindent}\textit{Local standardization}
\ForAll{$i \in \{1,\dots,N\}$ \textbf{in parallel}}
  \For{$j = 1,\dots,n_i$}
    \State \textbf{Client $i$:} $Y_{i,gj} \gets (Y_{i,gj}-\mu_g)/\sigma_g$
  \EndFor
\EndFor
\EndFor

\end{algorithmic}
\end{algorithm}

\begin{algorithm}[h]
\small
\caption{FPLIER training}
\label{alg:fed}
\begin{algorithmic}[1]
\Require Client training matrices $\{Y_i\}_{i=1}^N$; gene set matrix $C$; maximum number of iterations $T$; tolerance $\tau$
\Ensure PLIER model $(C,Z,U,\lambda_2)$

\Statex \hspace{-\algorithmicindent}\textit{Initialization}
\ForAll{$i \in \{1,\dots,N\}$ \textbf{in parallel}}
  \State \textbf{Client $i$:} Compute $Y_i Y_i^\top$
\EndFor
\State \textbf{Consortium:} Compute $Y Y^\top \gets \sum_{i=1}^N Y_i Y_i^\top$ via SecAgg+
\State \textbf{Server:} Compute randomized SVD $Y Y^\top \simeq V D^2 V^\top$
\State \textbf{Server:} Select $k$ and $\lambda_2$ via an elbow heuristic on $D$
\State \textbf{Server:} Communicate $k$, $\lambda_2$ to clients
\State \textbf{Server:} Set $\lambda_1 \gets \lambda_2/2$
\State \textbf{Server:} Initialize $\lambda_3$ to a small positive value
\ForAll{$i \in \{1,\dots,N\}$ \textbf{in parallel}}
  \State \textbf{Client $i$:} Let $n_i \gets$ number of columns of $Y_i$
  \State \textbf{Client $i$:} Compute local randomized SVD $Y_i = V_i D_i W_i^\top$
  \State \textbf{Client $i$:} Set $k_i \gets \min(k, n_i)$
\State \textbf{Client $i$:} Initialize $B_i^{(0)} \gets 0_{k \times n_i}$
\State \textbf{Client $i$:} Set the first $k_i$ rows of $B_i^{(0)}$ to $(W_{i,[:,1:k_i]} D_{i,1:k_i})^\top$
\EndFor

\Statex \hspace{-\algorithmicindent}\textit{Alternating updates}
\For{$\ell = 0,1,\dots,T-1$}

\Statex \hspace{\algorithmicindent}\textit{$Z$ update (federated, consortium)}

\ForAll{$i \in \{1,\dots,N\}$ \textbf{in parallel}}
      \State \textbf{Client $i$:} Compute $P_i \gets Y_i {B_i^{(\ell)}}^\top$
      \State \textbf{Client $i$:} Compute $Q_i \gets B_i^{(\ell)} {B_i^{(\ell)}}^\top$
\EndFor

  \State \textbf{Consortium:} Compute $P \gets \sum_{i=1}^N P_i$ via SecAgg+
  \State \textbf{Consortium:} Compute $Q \gets \sum_{i=1}^N Q_i$ via SecAgg+
  \State \textbf{Server:} $Z^{(\ell+1)} \gets \big(P + \lambda_1 C U^{(\ell)}\big)\big(Q + \lambda_1 I\big)^{-1}$
  \State \textbf{Server:} $Z^{(\ell+1)}[Z^{(\ell+1)}<0] \gets 0$
  \State \textbf{Server:} Broadcast $Z^{(\ell+1)}$ to all clients

  \Statex \hspace{\algorithmicindent}\textit{$B$ update (local, clients)}

\ForAll{$i \in \{1,\dots,N\}$ \textbf{in parallel}}
    \State \textbf{Client $i$:} $B_i^{(\ell+1)} \gets
      \big({Z^{(\ell+1)}}^\top Z^{(\ell+1)} + \lambda_2 I\big)^{-1}
      {Z^{(\ell+1)}}^\top Y_i$
      \label{alg:b_update}
\State Compute $a_i^{(\ell)} = \|B_i^{(\ell+1)} - B_i^{(\ell)}\|_F^2$
\State Compute $b_i^{(\ell)} = \|B_i^{(\ell+1)}\|_F^2$
\State Send $a_i^{(\ell)}, b_i^{(\ell)}$ to server
    \EndFor

    \Statex \hspace{\algorithmicindent}\textit{$U$ update (central, server)}
    
  \State \textbf{Server:} $U^{(\ell+1)} \gets \arg\min_U \|Z^{(\ell+1)} - C U\|_F^2 + \lambda_3 \|U\|_1$

  \State \textbf{Server:} Update $\lambda_3$ to control  $U^{(\ell+1)}$ sparsity (optional)

    \Statex \hspace{\algorithmicindent}\textit{Early stop}
\State \textbf{Server:} Stop if $\sum_{i=1}^N a_i^{(\ell)} / \sum_{i=1}^N b_i^{(\ell)} < \tau$

\EndFor

 \end{algorithmic}
\end{algorithm}

Before federated training, the expression matrices are standardized gene-wise across the consortium. In standard PLIER workflows, this is typically done by applying a global $z$-score normalization to the expression matrix. In the federated setting, the same transformation can be carried out without sharing individual samples by aggregating gene-level summary statistics.

Let $Y_i \in \mathbb{R}^{p \times n_i}$ denote the training matrix held by client $i$. For each gene $g$, client $i$ computes the number of local samples, the sum of expression values across its local samples, and the corresponding sum of squared expression values. These quantities are then securely aggregated across clients, allowing the server to compute the global gene-wise mean $\mu_g$ and standard deviation $\sigma_g$. The resulting statistics are broadcast to the clients, which then apply the standardization locally. Holdout samples, when used for the optional privacy analysis, are transformed using the same training-derived global statistics. Algorithm~\ref{alg:fed_standardization} summarizes this procedure.

\subsection{Federated PLIER training}
\label{appendix:training}

This section describes how the centralized PLIER training procedure can be reformulated in a federated setting. As noted earlier, the key observation is that both the initialization and the iterative updates depend on the data only through a small set of matrix products, which can be computed securely from client-level contributions. The overall procedure is summarized in Algorithm~\ref{alg:fed}.

Let $Y_i \in \mathbb{R}^{p \times n_i}$ denote the local standardized training matrix held by client $i$. 
Initialization of the PLIER model requires the covariance-like matrix
\[
Y Y^\top = \sum_{i=1}^{N} Y_i Y_i^\top ,
\]
which can be obtained by having each client compute its local contribution $Y_i Y_i^\top$ and submit it through secure aggregation. 
The server then computes an SVD of the aggregated matrix, from which the latent dimensionality $k$ and the regularization parameter $\lambda_2$ are selected. Following the standard PLIER procedure, $\lambda_1$ is set to $\lambda_2/2$.

PLIER also requires an initial value for the latent representation matrix $B$. In our implementation, the client-specific matrices $B_i^{(0)}$ used in the first training iteration are initialized locally. This avoids requiring the server to share with the clients information derived from the SVD of $Y Y^\top$, thereby reducing the exposure of intermediate quantities that could otherwise increase the risk of membership inference attacks. As a result, FPLIER uses a different initialization than the standard centralized PLIER procedure.

However, as described below, the alternating updates themselves remain algebraically equivalent to those of centralized PLIER, up to minor numerical differences (e.g., floating-point effects introduced by the SecAgg+ protocol). In particular, if the same initialization were used, the sequence of updates would coincide with that of the centralized PLIER. In practice, the use of different initial conditions is not problematic: PLIER optimization has been shown to be stable with respect to initialization and to converge reliably, and the alternative decompositions obtained at convergence are functionally equivalent~\cite{mao2019pathway}.

After initialization, the algorithm alternates among updates of $Z$, $B$, and $U$, mirroring the centralized procedure. More specifically:
\begin{itemize}
    \item the update of $Z$ is performed in a federated manner and requires the aggregated matrix products
\[
Y B^\top = \sum_{i=1}^{N} Y_i B_i^\top,
\qquad
B B^\top = \sum_{i=1}^{N} B_i B_i^\top,
\]
which are computed via SecAgg+;
    \item each client updates its matrix $B_i$ locally using its own data $Y_i$ and the shared matrix $Z$;
    \item the server updates $U$ using the current global matrix $Z$ and the  prior matrix $C$.
\end{itemize}

\clearpage

The stopping criterion can also be written in distributed form. Since the global matrix $B$ is partitioned column-wise  across clients, the relative change in $B$ can be decomposed as
\[
\frac{\|B^{(\ell+1)}-B^{(\ell)}\|_F^2}{\|B^{(\ell+1)}\|_F^2}
=
\frac{\sum_{i=1}^N \|B_i^{(\ell+1)}-B_i^{(\ell)}\|_F^2}
{\sum_{i=1}^N \|B_i^{(\ell+1)}\|_F^2},
\]
which follows from the additivity of the squared Frobenius norm over column blocks. As a result, the stopping criterion can be evaluated in a distributed manner while still measuring global convergence, relying only on aggregated, non-sensitive scalar quantities derived from client-side updates.

\section{Computational complexity and memory analysis}
\label{appendix:complexity}

\begin{table*}[t]

\caption{Dominant computational and memory costs associated with the components in which FPLIER differs from centralized PLIER. Here, $N$ denotes the number of clients, $p$ the number of genes, $n$ the total number of samples, $n_i$ the number of samples held by client $i$, $k$ the latent dimensionality used during training, and $w$ the number of components retained during randomized SVD initialization. Initialization costs are incurred once, whereas iterative costs are reported per training iteration. Computational costs for client-side operations are reported per client, while consortium-level operations are reported as total system costs. Rows align centralized and federated counterparts when applicable; entries marked ``--'' denote operations introduced only by the federated formulation. Operations with identical asymptotic cost in both methods are included only when their distributed execution is relevant.}
\label{tab:computational_costs_comparison}
\centering
\small
\setlength{\tabcolsep}{4pt}
\begin{tabular}{p{2.9cm} p{2.4cm} p{1.85cm} p{3.3cm} p{2.5cm} p{2.2cm} p{1.45cm}}
\toprule
\multicolumn{3}{c}{\textbf{Centralized PLIER}} &
\multicolumn{4}{c}{\textbf{FPLIER}} \\
\cmidrule(r){1-3} \cmidrule(l){4-7}
\textbf{Operation} & \textbf{Comput. cost} & \textbf{Memory cost} &
\textbf{Operation} & \textbf{Comput. cost} & \textbf{Memory cost} & \textbf{Executor(s)} \\
\midrule

\multicolumn{7}{l}{\textit{Initialization phase}} \\
\addlinespace[2pt]

Random. SVD on $Y$
& $\mathcal{O}(pnw)$
& $\mathcal{O}(pw+nw)$
&
Random. SVD on $Y_i$
& $\mathcal{O}(pn_iw_i)$
& $\mathcal{O}(pw_i+n_i w_i)$
& Client $i$ \\

-- 
& --
& --
&
Multipl. $Y_iY_i^\top$
& $\mathcal{O}(p^2 n_i)$
& $\mathcal{O}(p^2)$
& Client $i$ \\

-- 
& -- 
& --
&
SecAgg+ of $YY^\top$
& $\mathcal{O}(p^2N)$ overall
& $\mathcal{O}(p^2)$ per party
& Consortium \\

--
& --
& --
&
Random. SVD on $YY^\top$
& $\mathcal{O}(p^2w)$
& $\mathcal{O}(pw)$
& Server \\

\midrule

\multicolumn{7}{l}{\textit{Iterative phase (costs are per iteration)}} \\
\addlinespace[2pt]

Multipl. $YB^\top$
& $\mathcal{O}(pnk)$
& $\mathcal{O}(pk)$
&
Multipl. $Y_iB_i^\top$
& $\mathcal{O}(pkn_i)$
& $\mathcal{O}(pk)$
& Client $i$ \\

-- 
& -- 
& --
&
SecAgg+ of $YB^\top$
& $\mathcal{O}(pkN)$ overall
& $\mathcal{O}(pk)$ per party
& Consortium \\

Multipl. $BB^\top$
& $\mathcal{O}(k^2 n)$
& $\mathcal{O}(k^2)$
&
Multipl. $B_iB_i^\top$
& $\mathcal{O}(k^2 n_i)$
& $\mathcal{O}(k^2)$
& Client $i$ \\

-- 
& -- 
& --
&
SecAgg+ of $BB^\top$
& $\mathcal{O}(k^2N)$ overall
& $\mathcal{O}(k^2)$ per party
& Consortium \\

$B$ update
& $\mathcal{O}(pk^2+k^3+pnk)$
& $\mathcal{O}(kn)$
&
$B_i$ update
& $\mathcal{O}(pk^2+k^3+pkn_i)$
& $\mathcal{O}(kn_i)$
& Client $i$ \\

\bottomrule
\end{tabular}

\end{table*}

In this appendix, we summarize the dominant computational, communication, and memory costs of FPLIER. Let $N$ denote the number of clients, $p$ the number of genes, $k$ the latent dimensionality, $m$ the number of gene sets, $n$ the total number of samples in training, and $n_i$ the number of samples held by client $i$.

Relative to centralized PLIER, the additional costs of FPLIER arise from three main sources: (i) the different size of the matrices involved in the linear-algebra operations, (ii) the computation of auxiliary matrix products required by the distributed formulation, and (iii) the use of secure aggregation (SecAgg+).

We first summarize the dominant costs of the generic operations that are most relevant to PLIER and FPLIER.

\begin{itemize}

\item \textbf{Matrix addition.}
Adding two matrices of size $a \times b$ requires $\mathcal{O}(ab)$ arithmetic operations and $\mathcal{O}(ab)$ memory to store the result, excluding in-place implementations or temporary workspace \cite{golub}.

\item \textbf{Matrix multiplication.}
Multiplying matrices of size $a \times b$ and $b \times c$ requires $\mathcal{O}(abc)$ arithmetic operations and $\mathcal{O}(ac)$ memory to store the resulting matrix, excluding temporary workspace \cite{golub}.

\item \textbf{Matrix inversion.}
Inverting a dense matrix of size $a \times a$, or equivalently solving a dense linear system of the same dimension, requires $\mathcal{O}(a^3)$ arithmetic operations and $\mathcal{O}(a^2)$ memory \cite{golub}.

\item \textbf{Randomized SVD.}
For a matrix of size $a \times b$, computing a rank-$w$ randomized SVD, where $w$ is the number of retained singular components, requires leading computational cost $\mathcal{O}(abw)$ and dominant memory cost $\mathcal{O}(aw+bw)$ \cite{halko2011finding}.

\item \textbf{Secure aggregation (SecAgg+).}
Since Flower's SecAgg+ operates on vectorized client updates, the dominant payload-dependent overhead for large matrices is linear in both the number of clients and the number of transmitted entries. Thus, securely aggregating an object of dimension $a \times b$ across $N$ clients incurs a communication and aggregation cost of $\mathcal{O}(Nab)$, with $\mathcal{O}(ab)$ working memory at each participating party. Protocol setup costs, including key exchange and dropout recovery, are lower-order terms for the large matrix payloads considered here~\cite{li21, bonawitz2017practical}.

\end{itemize}

In the following, we discuss the main differences between centralized PLIER and FPLIER, summarized in Table~\ref{tab:computational_costs_comparison}. Client-side operations are performed in parallel across participants; therefore, computational costs for client-side operations are reported per client, whereas costs for consortium-level operations (e.g., secure aggregation) are reported as total system costs. Memory costs are always reported per participating party (clients and server).
Our focus is on the additional overhead introduced by the federated formulation, with the goal of clarifying how computational, communication, and memory costs are distributed across clients and the server.

For the iterative phase, costs are reported per iteration. The number of iterations is upper-bounded (at most 350), and in practice convergence is typically achieved earlier. Therefore, the number of iterations can be treated as $\mathcal{O}(1)$ and does not scale with the problem dimensions.

\subsection{Initialization}

The main difference in the initialization phase arises from the fact that centralized PLIER estimates the hyperparameters and latent dimensionality from a singular value decomposition (SVD) of $Y$, whereas FPLIER does so from the aggregated covariance-like matrix $YY^\top$. This requires two additional steps that are absent in the centralized setting: each client first computes its local matrix $Y_iY_i^\top$, and then these local contributions are combined via secure aggregation.

Forming $Y_iY_i^\top$ locally requires $\mathcal{O}(p^2 n_i)$ arithmetic operations and $\mathcal{O}(p^2)$ memory at client $i$. Secure aggregation of the resulting matrices across $N$ clients incurs a dominant communication and aggregation cost of $\mathcal{O}(Np^2)$ and requires $\mathcal{O}(p^2)$ working memory at both the server and each participating client.

In addition to these federated-only steps, SVD-based initialization is applied to different matrices in the two methods. Centralized PLIER performs randomized SVD directly on the pooled expression matrix $Y \in \mathbb{R}^{p \times n}$. In FPLIER, the server applies SVD to the aggregated covariance-like matrix $YY^\top \in \mathbb{R}^{p \times p}$ to determine $k$ and $\lambda_2$, while each client additionally performs a local SVD of $Y_i \in \mathbb{R}^{p \times n_i}$ to initialize $B_i$.

Using the generic randomized SVD cost summarized above with target rank $w$, centralized decomposition of $Y$ costs $\mathcal{O}(pnw)$ arithmetic operations with dominant memory $\mathcal{O}(pw+nw)$. In FPLIER, the server-side decomposition of $YY^\top$ costs $\mathcal{O}(p^2w)$ arithmetic operations, in addition to the $\mathcal{O}(p^2n_i)$ client-side cost of forming each $Y_iY_i^\top$ and the $\mathcal{O}(Np^2)$ secure-aggregation payload cost. Each client additionally incurs a local SVD cost of $\mathcal{O}(pn_iw_i)$, with memory $\mathcal{O}(pw_i+n_iw_i)$, where $w_i$ is the number of retained local components, typically $w_i \leq k$.

Consequently, centralized initialization scales jointly with the number of genes and samples, whereas FPLIER introduces a substantial $p^2$-scale initialization step through the formation, aggregation, storage, and decomposition of $YY^\top$. This cost is most problematic when $p$ is large, especially when $p \gg n$. It is more moderate only when dimensionality reduction makes $p$ small, or when $p$ is not substantially larger than the relevant sample sizes. From a memory perspective, the dominant FPLIER initialization requirement is storage of the dense $p \times p$ aggregated matrix.

\subsection{Iterations}

During the iterative phase, the main difference between centralized PLIER and FPLIER lies in the computation of the matrix products entering the $Z$ update. In centralized PLIER, these quantities are formed directly from the pooled matrices as $YB^\top$ and $BB^\top$. In FPLIER, the same quantities are instead obtained by first computing the corresponding client-level products $Y_iB_i^\top$ and $B_iB_i^\top$, and then aggregating them across clients through SecAgg+:
\[
YB^\top = \sum_{i=1}^N Y_iB_i^\top,
\qquad
BB^\top = \sum_{i=1}^N B_iB_i^\top.
\]

For client $i$, forming $Y_iB_i^\top$ requires $\mathcal{O}(pkn_i)$ arithmetic operations and $\mathcal{O}(pk)$ memory, while forming $B_iB_i^\top$ requires $\mathcal{O}(k^2n_i)$ arithmetic operations and $\mathcal{O}(k^2)$ memory. Relative to centralized execution, the additional overhead therefore arises from secure aggregation of the resulting $p \times k$ and $k \times k$ matrices, with dominant communication and aggregation costs of $\mathcal{O}(Npk)$ and $\mathcal{O}(Nk^2)$, respectively, and working memory $\mathcal{O}(pk)$ and $\mathcal{O}(k^2)$ at each participating party.

The update of $B$ is algebraically unchanged, but in FPLIER it is performed locally on each client using $Y_i$ rather than the full matrix $Y$. In centralized PLIER, the update
\[
B \gets (Z^\top Z + \lambda_2 I)^{-1} Z^\top Y
\]
consists of five dominant steps: forming the matrix $Z^\top Z \in \mathbb{R}^{k \times k}$ with cost $\mathcal{O}(pk^2)$, adding the term $\lambda_2 I$ with cost $\mathcal{O}(k^2)$, inverting the resulting matrix with cost $\mathcal{O}(k^3)$, computing $Z^\top Y$ with cost $\mathcal{O}(pkn)$, and multiplying the inverse by $Z^\top Y$ with cost $\mathcal{O}(k^2 n)$. Therefore, the total computational cost is
\[
\mathcal{O}(pk^2 + k^3 + pkn + k^2 n),
\]
since the $\mathcal{O}(k^2)$ addition is lower order. In typical settings where $p \gg k$, the term $\mathcal{O}(pkn)$ dominates $\mathcal{O}(k^2 n)$, yielding the simplified expression
\[
\mathcal{O}(pk^2 + k^3 + pkn).
\]
The memory cost is $\mathcal{O}(kn)$ to store the updated matrix $B$.

In FPLIER, client $i$ performs the analogous local update
\[
B_i \gets (Z^\top Z + \lambda_2 I)^{-1} Z^\top Y_i \,.
\]
The computational cost is therefore
\[
\mathcal{O}(pk^2 + k^3 + pkn_i) \,,
\]
since the last operation is data-dependent; the memory cost to store $B_i$ is also data-dependent and is $\mathcal{O}(kn_i)$ for each client.

Consequently, centralized iterations scale with pooled matrix products across all $n$ samples, whereas FPLIER replaces these operations with client-level products, followed by secure aggregation. Since the same computations are distributed across clients, the additional per-iteration overhead is generally modest and arises primarily from communication during aggregation. From a memory perspective, the dominant temporary storage in the iterative phase is $\mathcal{O}(pk+k^2)$, substantially smaller than the $\mathcal{O}(p^2)$ memory required during initialization.

\section{SVD, eigendecomposition, and PCA information}
\label{appendix:privacy}

This appendix clarifies the relationship among the singular value decomposition (SVD) of $Y$, the eigendecomposition of $YY^\top$, and the principal component information exposed by $YY^\top$.

Let
\[
Y = V D W^\top
\]
denote the SVD of the training expression matrix $Y$, where $D$ contains the singular values of $Y$. Then
\[
YY^\top = V D D^\top V^\top = V \Lambda V^\top,
\]
where $\Lambda = D D^\top$ contains the eigenvalues of $YY^\top$.

Thus, the eigendecomposition of $Y Y^\top$ recovers the left singular vectors of $Y$. Since $Y Y^\top$ is symmetric positive semidefinite, its singular values equal its eigenvalues, and one valid SVD is obtained by taking the eigenvectors as both left and right singular vectors. Consequently, efficient truncated or randomized SVD methods can be applied directly to $Y Y^\top$ to obtain approximate leading components without computing the full decomposition, which is advantageous in large-scale settings.

The eigenvalues in $\Lambda$ are the squared singular values of $Y$, so the singular values of $Y$ are obtained as the square roots of the eigenvalues of $Y Y^\top$. Equivalently, this decomposition reveals the principal component directions and their associated variances (depending on normalization).

\end{document}